# DNA Methylation Data to Predict Suicidal and Non-Suicidal Deaths: A Machine Learning Approach

Rifat Zahan
Department of Computer Science
University of Saskatchewan
Saskatoon, Saskatchewan, Canada
Email: rifat.zahan@usask.ca

Ian McQuillan
Department of Computer Science
University of Saskatchewan
Saskatoon, Saskatchewan, Canada
Email: mcquillan@cs.usask.ca

Nathaniel D. Osgood
Department of Computer Science
University of Saskatchewan
Saskatoon, Saskatchewan, Canada
Email: nathaniel.osgood@usask.ca

*Abstract*—Suicide is one of the leading causes of death. Major Depressive Disorder (MDD) is one of the risk factors for committing suicide. Epigenetic data may help to distinguish suicidal and non-suicidal deaths. In particular, DNA methylation is a process involving a chemical modification on DNA which can change gene activity without changing the sequence. It has successfully been used for monitoring suicide progression and prediction. The objective of this study is to predict suicidal and non-suicidal deaths from DNA methylation data using a modern machine learning algorithm. We used support-vector machines to classify existing secondary data consisting of normalized values of methylated DNA probe intensities from tissues of two cortical brain regions (Brodmann Area 11 [BA11] and Brodmann Area 25 [BA25]) to distinguish 20 suicide cases (who died following a major depressive disorder) from 20 control cases (non-psychiatric sudden death). Prior to classification, we employed Principal component analysis (PCA) and t-distributed Stochastic Neighbor Embedding (t-SNE) to reduce the dimension of the data. In comparison to PCA, the modern data visualization method t-SNE performs better in dimensionality reduction. t-SNE accounts for the possible non-linear patterns in low-dimensional data. By contrast, PCA failed to group the cases according to different types of deaths. We applied four-fold cross-validation in which the resulting output from t-SNE was used as training data for the Support Vector Machine (SVM). For BA25, two-dimensional t-SNE better classified the suicidal vs. non-suicidal deaths compared to three-dimensional t-SNE. The Receiver Operating Characteristic Curve (ROC) for the classification exhibited an 80% Area Under the Curve (AUC) for BA25 data and 100% AUC for BA11. Despite the use of cross-validation, the nominally perfect prediction of suicidal deaths for BA11 data suggests possible over-fitting of the model. The study also may have suffered from 'spectrum bias' since the individuals were only studied from two extreme scenarios. This research constitutes a baseline study for classifying suicidal and non-suicidal deaths from DNA methylation data. Future studies with larger sample size, while possibly incorporating methylation data from living individuals, may reduce the bias and improve the accuracy of the results.

## I. INTRODUCTION

According to the World Health Organization (WHO), each year about 800,000 suicides occur worldwide, and there are a countless number of people who attempt suicide [1]. About 1.4% of all deaths are related to suicide [1]. Suicide is one of the leading causes of deaths in the United States [2] and is associated with significant social, economic and health system cost [3]. Mental health disorders are considered to be one of the significant risk factors for people committing suicide in developed countries [4]. Indeed, approximately, 90% of those who completed suicide had some mental health disorder in their lifetime [5]. One of the suicide prevention strategies proposed by the National Action Alliance for Suicide Prevention is to identify biomarkers that may help to identify people who are at high risk of suicide [6]. The epigenetic effects associated with Major Depressive Disorder (MDD) related suicides are not particularly well-studied [4]. In recent years, DNA methylation data has been successfully used to identify cancer at an early stage and to monitor subsequent tumor progression of the individuals throughout their treatments [7]. DNA methylation refers to the addition of methyl group ($CH_3$) to a DNA molecule [8], which controls gene expression [9]. One of the crucial objectives of the study of DNA methylation is to find the DNA loci where methylation is related to clinical, social, demographic factors and disease status [10]. In human beings, methylation is usually found in locations of DNA called cytosine-guanine dinucleotide (CpG or CG) sites [11, 12]. Researchers have found Differentially Methylated Regions (DMR) that are hypomethylated in the cortex of brain regions of suicidal people [4]. DNA methylation data has also been successfully used in the epidemiology of psychiatric disorder and suicidal behavior [4, 13, 14]. Around early 1997, in the work of Harris and Barraclough, the authors did not report any causal association between DNA methylation and suicide, but the authors reported that the number of cytosine sites increased dramatically in the brains of suicide victims compared to non-psychiatric non-suicidal brains [5]. Altered DNA methylation and associated co-methylation are highly associated with increased rate of suicide [4].

Only a few studies have been conducted so far to identify possible genomic biomarkers for the prediction of suicide with high accuracy [14]. Kaminsky et al. generated DNA methylation values obtained from saliva [15]. The authors used a multiple linear regression model to predict suicidal behavior while controlling for age, sex, ethnicity and cytosine-guanine dinucleotide (CpG) content [15]. The authors reported that although the model was useful for predicting suicidal behavior (suicidal ideation and suicide attempts), it produced poor accuracy, yielding an AUC was 55% for suicidal ideation and 15% for suicide attempt [15].

The relatively limited research in the prediction of suicide mortality from epigenetic data motivates this research. The objective of this study is to identify a model to classify suicide-related deaths and non-suicide related deaths from DNA methylation data using an appropriate machine learning algorithm.

## II. DATA DESCRIPTION

### A. Data Source

For this study, secondary data on DNA methylation was obtained from the National Center for Biotechnology Information (NCBI), GEO Accession: GSE88890 [4]. Murphy et al. [4] collected the data primarily from Douglas-Bell Canada Brain Bank (DBCBB) [16]. The authors collected brain tissue from 20 major depressive disordered suicide cases and 20 non-psychiatric sudden (NPS) death controls [4]. Tissues were obtained from two cortical brain regions: Brodmann Area 11 (BA11) and Brodmann Area 25 (BA25) [4], and genomic DNA was further extracted from the tissues.

Samples were processed and assessed using Illumina Infinium HumanMethylation450K BeadChip (Illumina) [4]. The raw signal of each probe was extracted using Illumina Genome Studio Software [4]. CpG contents of normalized $\beta$ values were used for further analysis, where

$$\beta = \frac{\text{MPI}}{\text{MPI} + \text{UPI}}, \quad (1)$$

where, MPI = Methylated Probe Intensities and UPI = Unmethylated Probe Intensities [4]. The authors reported 416,876 probes from 75 samples. Of those, 35 samples were for brain region BA25 (17 suicide cases and 18 non-suicide cases), and 40 matched samples of BA11 were taken from the same individuals (20 suicide cases and 20 non-suicide cases) [4]. Finally, in the data, there were a total of 327,616 normalized CpG contents for both of the cortical brain regions.

### B. Data Preparation for Analysis

Given such a large number of CpG contents, it becomes computationally inconvenient to analyze the data. Therefore, we have randomly selected 15,000 CpG contents out of 327,616 CpG contents for further analysis. Since we are using a machine learning algorithm for prediction of the suicidal and non-suicidal deaths, 75% of the data (*30 individuals for BA11 and 27 individuals for BA25*) are retained for training the model, which will be tested in the remaining 25% of the data (*10 individuals for BA11 and 8 individuals for BA25*), representing a 4-fold cross-validation (CV).

## III. METHODOLOGY

### A. Dimensionality Reduction

We are aiming to predict (or classify) the suicide-related deaths and non-suicidal deaths given the normalized $\beta$-values of CpG contents derived from DNA methylation data. Dimensionality reduction is an essential step in classification if the number of observations is less than the number of features [17]. Since we have high-dimensional data due to such a high number of features, we need to reduce the dimension in a way retains essential aspects of the data [18]. We considered two popular methods for dimensionality reduction: Principal Component Analysis (PCA) and t-distributed Stochastic Neighbor Embedding (t-SNE), which are described below:

*1) **Principal Component Analysis (PCA)***: PCA is a *linear algorithm* that *maximizes* the *variance* of the projection of the high dimensional data into low dimension space while retain the original information of the data [17]. PCA captures *global structure* (i.e., clusters at multiple scales) of the data [18, 19].

*2) **t-distributed Stochastic Neighbor Embedding (t-SNE)***: t-SNE is based on SNE, which was presented by Hinton and Roweis [20]. SNE converts high dimensional Euclidean distances (between points in the data) into conditional probabilities resembling similarities [18]. SNE suffers from the 'crowding problem,' since the cost function is difficult to optimize, hindering the quality of the visualization [18]. This problem can be minimized by using t-SNE, where the cost function involves a simpler gradient, using the Student's t-distribution instead of a Gaussian distribution [18]. This approach simplifies the calculation of distances between similar points. t-SNE is a *non-linear algorithm* that captures the *global*, as well as the *local* structure of high-dimensional data [18]. t-SNE is one of the best methods for use in dimensionality reduction [21]. t-SNE projects the neighborhood structure of the data in a way that points that are *similar* will *remain closer* in the space [18]. t-SNE is *not* a *clustering algorithm* since the input features are no longer identifiable. t-SNE suffers from time and space complexity since it faces the challenge of working with sometimes more than 10,000 features. [18].

The *output* from low dimensional features can be used for training the classifier.

### B. Classifier: Support Vector Machine (SVM)

In machine learning, classification refers to categorizing a new set of data based on feature(s). SVM is one of the most powerful tools to gain popularity in the area of machine learning and bioinformatics [17]. SVM is a classifier that finds a *hyperplane* which separates classes reasonably well [17]. It has three main tuning parameters:

(i) Kernel: The two groups may not be classified linearly [17]. The kernel can work to create a non-linear hyperplane to separate the classes. The kernels can be linear, polynomial, radial basis and sigmoid [22]. The kernel is useful in high dimensional settings [23].

(ii) Gamma ($\gamma$): kernel coefficient (radial-basis specific) [22].

(iii) Cost (*C*): penalty parameter. The cost parameter interacts with the gamma coefficient [23]. The higher the *γ*, the lower the *C* [23].

One interesting property of SVMs is that they are robust to outliers [22]. Therefore, SVM can plot a hyperplane in a classification setting where outliers are present.

Statistical software package R [24] is used to conduct all the analysis.

## IV. RESULTS

As mentioned in Section III-A, the dimensionality of the data needs to be reduced while retaining essential features of the high-dimensional data. In this section, both PCA and t-SNE are investigated to reduce both the BA11 and BA25 data into two-dimensional and three-dimensional settings. We have also compared between the results derived from these two methods (i.e., PCA and t-SNE).

From Figure 1(a), we see that t-SNE reduces the high-dimensional data into a low-dimension (2D) in a way that makes it easier to distinguish the two groups of deaths in the data of both cortical brain regions. Although PCA reduces the dimensionality, it offers limited differentiation between the suicidal and non-suicidal classes. Therefore, while reducing the high-dimensional data into a low-dimensional space, we judged it better to use t-SNE rather than PCA (see Figure 1(a) and 1(b)).

We have also examined the results of reducing high-dimensional data into three-dimensional space using PCA and t-SNE. From Figure 2, we can see that although PCA reduced the dimensionality, still it lacks the ability to separate cases according to the type of death experienced by the donors. However, t-SNE strongly differentiates between suicidal and non-suicidal deaths for both of the cortical brain regions. For BA11 (see Figure 2(a)), t-SNE separates the groups (i.e., types of deaths) into two distinct groups. By contrast, for BA25 (see Figure 2(b)), t-SNE generates some outliers by assigning non-suicidal sudden deaths into upper right and lower left corners of the plot. Such variation in the visual representation justifies that the use of an SVM since it works towards such outliers by creating a non-linear hyperplane. Such a hyperplane can accurately classify suicidal and non-suicidal deaths.

On the basis of the investigation results above, we selected t-SNE as the input for SVM (both two-dimensional and three-dimensional scenarios). Since the total number of observations in the training set is 30 individuals for BA11 and 27 individuals for BA25 data, therefore, we cannot go beyond using more than three features as input in SVM.

As noted above, SVM is highly sensitive to certain parameters, including the cost parameter (*C*) and gamma (*γ*). In order to identify the model that shows the best performance, we sought to tune those parameters. For tuning the model, we used two-fold cross-validation. When tuning the parameters several times, the most favourable model was obtained when *C* = 0.1 and *γ* = 0.5. This means that such a combination of SVM parameters returned models with the minimum encountered error estimates. However, for BA11 data, the most favorable model was obtained using the *polynomial* kernel, whereas, for BA25, we need the *radial basis* kernel. The reason for such difference in kernel parameter is because of the outliers in BA25 data, which are not present in BA11 data. A *polynomial* kernel is sufficient to capture the non-linear hyperplane in the BA11 SVM model.

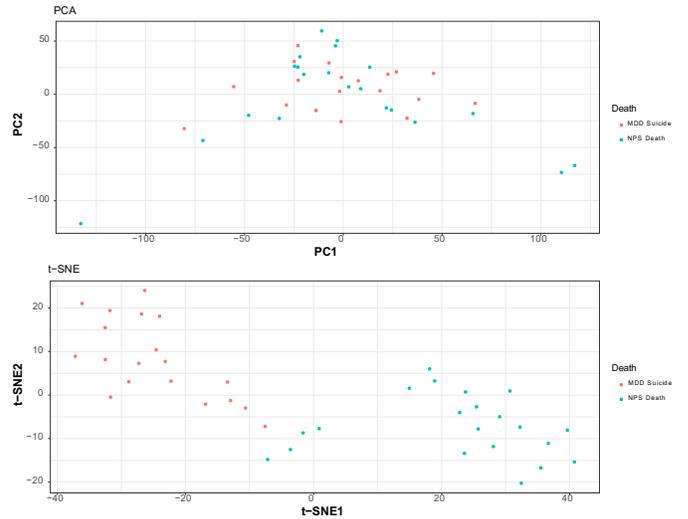

(a) Two Dimensional PCA and t-SNE for BA11

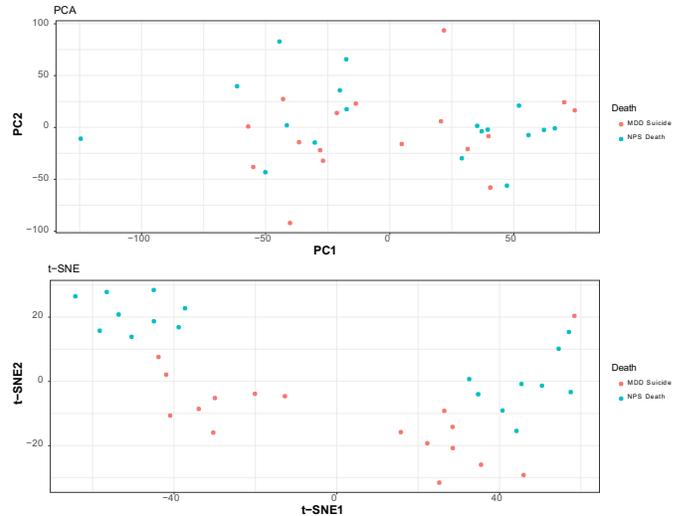

(b) Two Dimensional PCA and t-SNE for BA25

Fig. 1: Dimensionality reduction using two-dimensional PCA and t-SNE.

The resulting hyperplane from two-dimensional t-SNE is presented in Figure 3. As can be recognized from these two figures, the hyperplane is able to differentiate the classes of suicidal and non-suicidal deaths reasonably well. There are few misclassifications of the type of deaths for BA11 data (see Figure 3(a)). In the classification based on BA25 data (see Figure 3(b)) we see that the hyperplane classified the type of deaths in two different regions. Such classification shows the robustness of SVM towards outliers.

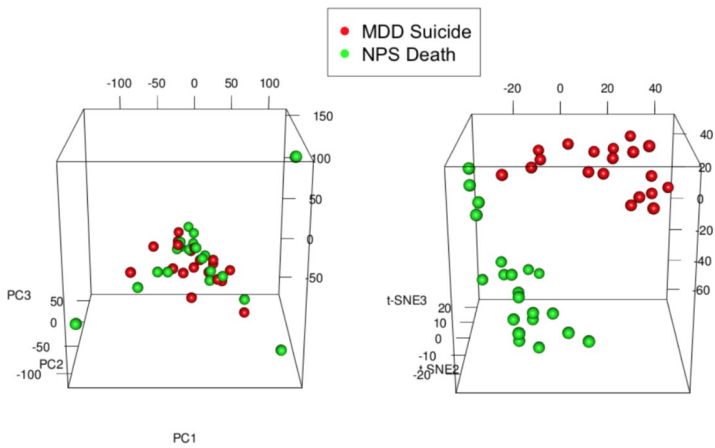

(a) Three Dimensional PCA and t-SNE for BA11

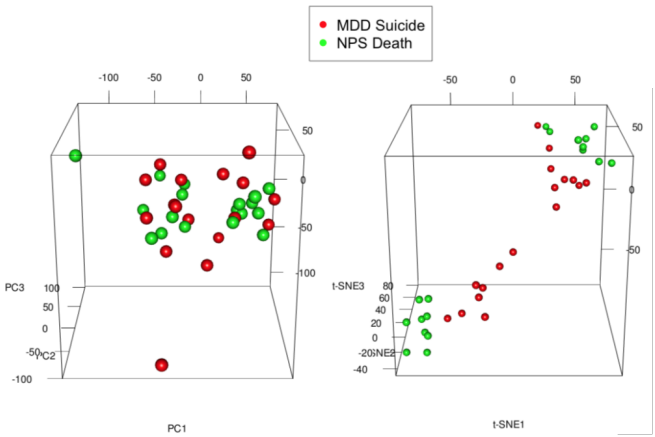

(b) Three Dimensional PCA and t-SNE for BA25

Fig. 2: Dimensionality reduction using three-dimensional PCA and t-SNE.

We further classified the classes of deaths using three-dimensional t-SNE for both BA11 and BA25 data-sets. The resulting hyperplanes are presented in Figure 4. We can see from one angle that for BA11 model, there are some misclassifications of non-suicidal sudden deaths resulting from the hyperplane (see Figure 4(a)), even though the hyperplane distinguished the classes reasonably well. The fitted hyper-planes from BA25 data shows that it differentiates the groups relatively well, even though they lie in two different regions in the plot.

*A. Cross-Validation*

We have found favorable models for both BA11 and BA25 data for the prediction (or classification) of the MDD Suicide cases and Non-Psychiatric Sudden (NPS) Deaths. However, we have fitted models for both two-dimensional and three-dimensional t-SNE. Therefore, to find the best-suited model for such scenario, we need to perform cross-validation (CV) to test the model performance. As mentioned in Section II-B, we have retained 25% of the data from both of the brain cortical regions to test our trained SVM model. Therefore, the trained model based on 75% of the data is tested on the remaining 25% of the data. The ROC curves for both the data using different dimensions are presented in Figure 5.

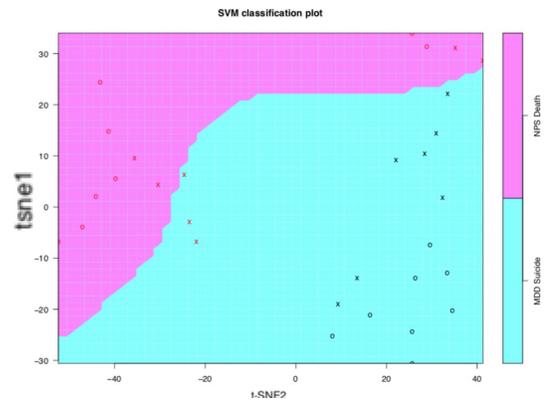

(a) SVM on BA11

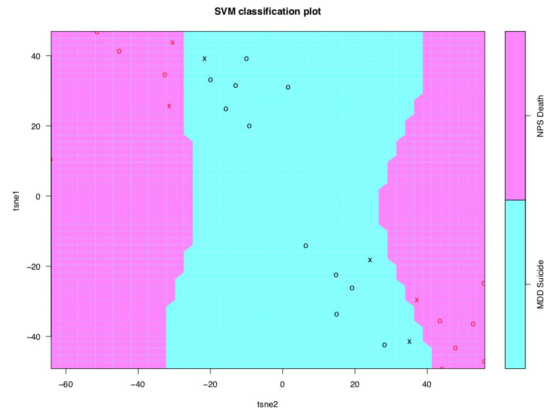

(b) SVM on BA25

Fig. 3: Fitted hyper-plane of SVM from two-dimensional t-SNE.

Figure 5 shows that for two-dimension t-SNE, the resulting hyperplane generated an Area Under the Curve (AUC) of 95%. By contrast, AUC was 100% for three-dimensional t-SNE. Such high AUC indicates that the model may have over-fitted the data. As for BA25 (see Figure 5(b)), the two-dimensional t-SNE resulted in higher accuracy compared to three-dimensional t-SNE. In Statistics, researchers use a 'rule of thumb' that there should be at least 10 observations for each covariate (or features) [25, 26, 27]. In BA25 data, we have 27 observations for training the model. This suggests that limited benefit would be obtained from considering more than two dimensions.

The results, demonstrate that two-dimensional t-SNE is useful to train the SVM model for BA25 and can be used for high accuracy classification of suicidal and non-suicidal deaths. Although the two-dimensional and three-dimensional t-SNE from BA11 data produced an extremely high accuracy classifier by training the SVM, this prediction may not be useful for predicting suicidal and non-suicidal deaths. There-fore, results from this study based on BA11 data may not be recommended for another population for prediction purpose.

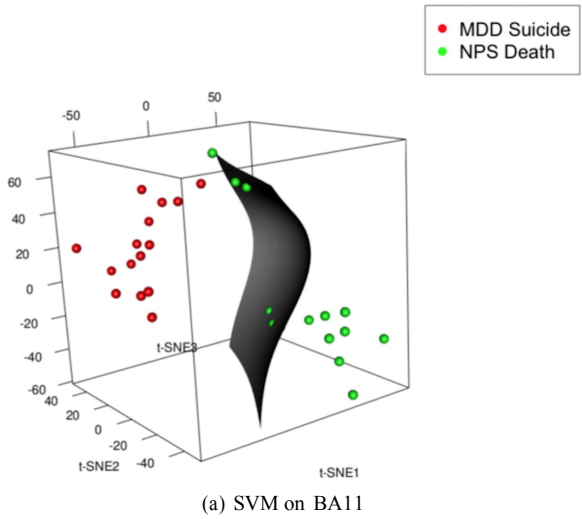

(a) SVM on BA11

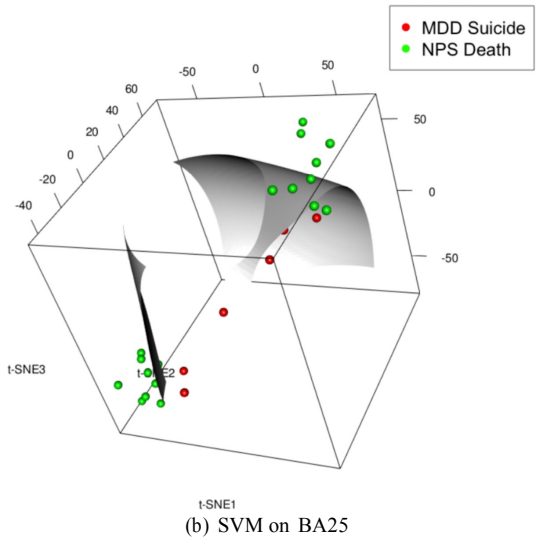

(b) SVM on BA25

Fig. 4: Fitted hyper-plane of SVM from two-dimensional t-SNE.

## V. CONCLUSION

The interest in epigenetic effects in the study of suicide is increasing worldwide to understand the etiology of suicidal behavior and suicide progression, since this area is not yet well-studied. In this study, we have used DNA methylation in Illumina 450K data to classify the suicidal and non-suicidal deaths by employing a machine learning approach.

In comparison to PCA, the modern data visualization method t-SNE performs better in dimensionality reduction. t-SNE accounts for the possible non-linear pattern in low-dimensional data. By contrast, PCA failed to group the cases according to different types of deaths.

The resulting data reduced via t-SNE were used for further analysis to predict suicide-related deaths using SVM. SVM fit the hyperplane in the data reasonably well. Two-dimensional t-SNE performs better compared to three-dimensional t-SNE for BA25 data. Even though both two-dimensional and three-dimensional t-SNE fit the BA11 data reasonably well, a problem of over-fitting may have occurred in BA11 data as it can occur when using complex methods for training the model [28]. For example, usage of complex kernels rather than simple kernels may have caused complexity in the model.

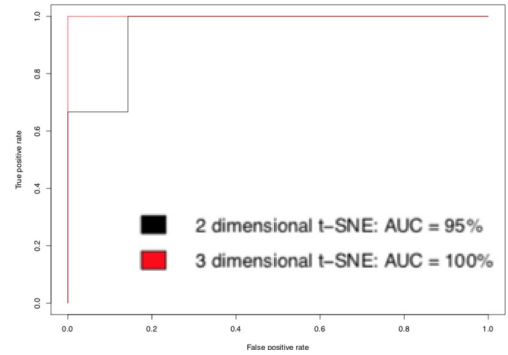

(a) ROC for BA11

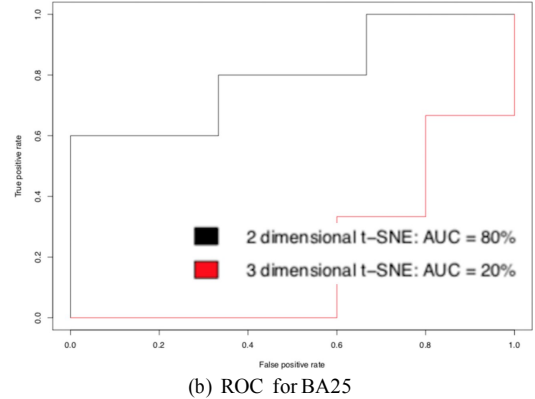

(b) ROC for BA25

Fig. 5: Area Under Curve (AUC) in Receiver Operating Characteristic (ROC) Curve for the final model derived from Support Vector Machine (SVM).

One of the drawbacks of this study is 'spectrum effect,' which is less appreciated in health-related research. We considered two extreme groups (suicides following mental disorder and non-suicidal deaths of healthy individuals) in the analysis. Such an extreme choice of the groups of study individuals are more likely to increase sensitivity and reduce false negatives [29]. This may be another reason for a perfect prediction for suicide based on BA11 data using three dimensions. Although SVM can perform well when the sample size is small [30], a large sample test set is required for a classifier with low error rate [31]. We had only a few observations for cross-validation. Therefore, caution should be exercised in classifying suicidal and non-suicidal deaths based on the models identified using this study. Increasing the sample size in test data is required before recommending the classifier resulting from this study for other populations. Since obtaining DNA methylation data is quite expensive and time-consuming; therefore, simulation study should be considered to obtain synthetic ground truth data. Such an approach will allow to train the model using synthetic data and test the model in the real data for cross-validation. Future study is needed to examine that if a simple choice for kernels in the SVM models is enough to train the classifier to avoid the problem of over-fitting. The 'spectrum effect' can be eliminated if data can be obtained from living individuals: obtaining DNA methylation data from saliva, blood or hair sample. This way, obtaining a sample from the continuum of suicide progression will be possible, which will be more useful to predict suicide and non-suicidal deaths.

This is a baseline study for classifying suicidal and non-suicidal deaths from DNA methylation data using a machine learning algorithm. However, overcoming the challenges encountered in this study may help the health-care providers:
(i) to act towards identifying individuals who are at high risk of suicide and (ii) to provide early level intervention to help prevent suicides and suicide attempts.